\documentstyle[multicol,prb,aps,graphicx,amssymb]{revtex}
\setkeys{Gin}{width=0.9\linewidth}

\begin{document}
\draft
\title{Diamagnetic response of cylindrical normal metal -- 
superconductor proximity structures with low
concentration of scattering centers}
\author{F.~Bernd~M{\"u}ller-Allinger and Ana Celia Mota}
\address{Laboratorium f{\"u}r Festk{\"o}rperphysik, Eidgen{\"o}ssische
         Technische Hochschule Z{\"u}rich, \\
         8093 Z{\"u}rich, Switzerland}
\author{Wolfgang Belzig}
\address{Institut f{\"u}r Theoretische Festk{\"o}rperphysik,
Universit{\"a}t Karlsruhe, D-76128 Karlsruhe, Germany}
\date{\today}
\maketitle

\begin{abstract}
We have investigated the diamagnetic response of composite NS 
proximity wires, consisting of a clean silver or copper coating, in 
good electrical contact to a superconducting niobium or tantalum core.  
The samples show strong induced diamagnetism in the normal layer, 
resulting in a nearly complete Meissner screening at low temperatures.  
The temperature dependence of the linear diamagnetic susceptibility 
data is successfully described by the quasiclassical Eilenberger 
theory including elastic scattering characterised by a mean free path 
$\ell$.  Using the mean free path as the only fit parameter we found 
values of $\ell$ in the range $0.1-1$ of the normal metal layer 
thickness $d_{\mathrm N}$, which are in rough agreement with the ones 
obtained from residual resistivity measurements.  The fits are 
satisfactory over the whole temperature range between 5\,mK and 7\,K 
for values of $d_{\mathrm N}$ varying between $1.6\,\mu\mathrm m$ and 
$30\,\mu\mathrm m$.  Although a finite mean free path is necessary to 
correctly describe the temperature dependence of the linear response 
diamagnetic susceptibility, the measured breakdown fields in the 
nonlinear regime follow the temperature and thickness dependence given 
by the clean limit theory.  However, there is a discrepancy in the 
absolute values.  We argue that in order to reach quantitative 
agreement one needs to take into account the mean free path from the 
fits of the linear response. [PACS numbers: 74.50.+r, 74.80.-g]
\end{abstract}

\begin{multicols}{2}
\narrowtext
\section{Introduction}
A normal metal (N) in good electrical contact with a superconductor 
(S) exhibits superconducting properties as the temperature is reduced.  
First experiments on the proximity effect were reported by R.~Holm and 
W.~Meissner\cite{holm}, who observed zero resistance between SNS 
pressed contacts.  Since then, many investigations on proximity 
effects have been carried out\cite{deutscher}.  Recently, it has 
received a revived interest\cite{review}.  Particularly, experiments 
on the magnetic response have demonstrated nontrivial screening 
properties, showing hysteretic magnetic breakdown at finite external 
fields\cite{mota82,bergmann87,expsetup} as well as a presently 
unexplained reentrant effect at low temperatures\cite{visaniprl}.  

First theoretical studies on the proximity effect were carried out by 
Cooper\cite{cooper61}.  For a diffusive proximity system the 
diamagnetic susceptibility in the linear regime was investigated by 
the Orsay Group on Superconductivity in the framework of the 
Ginzburg--Landau theory\cite{degennes}.  In this limit it was found 
that the induced diamagnetic susceptibility in N depends on 
temperature approximately as $\chi_{\mathrm N}\propto T^{-1/2}$.  This 
was confirmed experimentally by Oda and Nagano\cite{oda:80}.  In spite 
of that, the magnetic properties of proximity samples discussed in 
Refs.\onlinecite{mota82} and \onlinecite{expsetup} could not be explained 
by the Ginzburg--Landau theory, since they show a much stronger 
temperature dependence of the diamagnetic susceptibility.

The clean limit, which is defined for the elastic mean free path $\ell 
\to \infty$, was first studied by Zaikin\cite{zaikin} for a finite 
system of ideal geometry with the help of the quasiclassical theory.  
He predicted that, since in this case the current--field relation is 
completely nonlocal, the current in N is spatially constant and 
depends on the vector potential integrated over the whole normal 
metal.  As a consequence, the magnetic flux is screened linearly over 
the normal layer thickness $d_{\mathrm N}$ and for $T\to 0$ the 
susceptibility can reach only $3/4$ of that of a perfect diamagnet.  
For $T\gg T_{\mathrm A}$ he found $\chi_{\mathrm N}\propto 
exp(-2T/T_{\mathrm A})$, with the Andreev temperature $T_{\mathrm 
A}=\hbar v_{\mathrm F}/2\pi k_{\mathrm B} d_{\mathrm N}$.  This 
mesoscopic temperature scale originates from the energy of the lowest 
Andreev bound state\cite{andreev}.  In this ballistic case all 
properties of the proximity effect are determined by these bound 
states, which are coherent superpositions of electron and hole waves 
between consecutive Andreev and normal reflections.

Within the framework of the quasiclassical theory, Belzig~\textsl{et 
al.}\cite{belzig:96} studied recently the magnetic response of a 
proximity coupled NS sandwich in the two limits, clean and dirty.  
They obtained numerical solutions of the corresponding equations for a 
wide range of temperature, magnetic field and layer thickness.  
Furthermore, they tried a fit of the susceptibility data of one AgNb 
specimen with a relatively small mean free path (specimen 1AgNb in 
this paper) within their dirty limit results.  The fit was only 
successful at very low temperatures, but could not reproduce the high 
temperature data.  The diamagnetic screening of the dirty limit 
was too big as compared to the experiment.  Moreover, the temperature 
dependence of the susceptibility of NS proximity specimens with the 
longest mean free paths could not be explained within the clean limit 
theory.  This limit gives a completely different temperature 
dependence than the one observed in the experiments.  In addition, 
data at low temperatures reach almost perfect diamagnetic screening, 
whereas in the clean limit it should not exceed $75\,\%$ of full
screening.
To close the gap between these two limits the linear magnetic response 
for arbitrary impurity concentrations was theoretically studied in 
Ref.\onlinecite{belzig}.  We show in this paper that these 
results can satisfactorily describe the experimental data.

The theory of the nonlinear reponse was addressed in 
Refs.\onlinecite{degennes,belzig:96}, and 
\onlinecite{fauchere}.  A proximity NS sandwich with a finite normal 
layer thickness $d_{\mathrm N}$ at sufficiently low temperatures 
undergoes a first order phase transition both in the clean and in the 
dirty limit.  At a breakdown field a jump in the magnetization occurs.  
In the clean limit the temperature dependence of the breakdown field 
is $H_{\mathrm b}(T)\propto exp(-T/T_{\mathrm A})$. We show in this 
paper, that the clean limit result agrees well with our experiments on 
$H_{\mathrm b}$ (see also Ref.\onlinecite{expsetup}), which follow the 
same temperature dependence in relatively clean specimens 
($\ell/d_{\mathrm N}\approx 0.4-0.8$).  Also the experimentally found 
absolute values of the breakdown fields show a $1/d_{\mathrm 
N}$--dependence, which is in qualitative agreement with the clean 
limit theory\cite{fauchere}.  However, this is not the case for the 
shape of the magnetization curves and, as mentioned before, for the 
temperature dependence of the linear susceptibility.

We have fabricated a new set of CuNb and AgNb specimens similar to the 
ones reported in Ref.\onlinecite{expsetup}.  The specimens were 
produced with an optimized annealing procedure in order to achieve 
very high mean free paths.  The ratio of the thickness of the normal 
layer to the radius of the superconductor of the composite wires, 
$d_{\mathrm N}/r_{\mathrm S}$, was also varied to investigate 
comparable samples with normal layers reaching from almost flat to 
rather curved cylindrical geometry.  Thus the influence of the NS 
geometry on the proximity effect was investigated.  In this paper, we 
discuss these new samples together with older ones\cite{expsetup}, 
covering a wide range of parameters, and fit their diamagnetic 
response with the help of the quasiclassical theory in an intermediate 
impurity regime between the clean and dirty limit. 
Here, we will not consider the reentrant effect found by Mota and 
co--workers\cite{visaniprl} but only note that a 
recent theoretical study has addressed this new 
phenomenon\cite{bruder}. For the nonlinear 
response we compare the temperature dependent breakdown fields with 
the quasiclassical clean limit result.

This paper is organized as follows.  In Sec.~\ref{sec:theory} we 
describe some theoretical aspects of the magnetic response.  In 
Sec.~\ref{sec:setup} we describe the sample preparation and the 
measurement apparatus.  In Sec.~\ref{sec:linearresponse} the 
experimental results on the linear magnetic response are presented and 
the data are fitted using the theory from Sec.~\ref{sec:theory}.  The 
experimental results on the nonlinear response are addressed in 
Sec.~\ref{sec:nonlinear}.  Finally we draw some conclusions.

\section{Theoretical Aspects}
\label{sec:theory}
Within the quasiclassical theory of 
super\-conduc\-ti\-vi\-ty\cite{eilenberger,larkin,rainersauls} the 
magnetic response of a planar NS--proximity structure was investigated 
for arbitrary impurity concentrations by Belzig~\textsl{et 
al.}\cite{belzig}.  Below we summarize some basic results of 
this paper.  The theoretical system consists of a semi--infinite 
superconductor in perfect contact with a normal metal of thickness 
$d_{\mathrm N}$ and specular reflecting outer surface.  The pair 
potential $\Delta$ is taken to be constant in the superconductor and 
to be zero in the normal metal.  The mean free path $\ell$ and the 
Fermi velocity $v_{\mathrm F}$ are assumed to be the same throughout 
the system.

In Ref.\onlinecite{belzig} a variety of regimes are discussed, where 
the proximity effect is different from the previously studied clean 
and dirty limits.  The regimes differ by the relative magnitudes of 
the thermal length $\xi_T=\hbar v_{\mathrm F}/2\pi k_{\mathrm B} T$, 
the mean free path $\ell$, and the thickness $d_{\mathrm N}$, an 
additional relevant length scale, which has to be taken into account.  
In the case of a finite normal layer thickness $d_{\mathrm N}$, the 
clean and the dirty limit are restricted to much smaller parameter 
regions than previously believed.  The dirty limit holds for 
$\ell\ll\xi_{T},d_{\mathrm N}$ only, if also the mean free path 
$\ell\ll\xi_{0}$.  Here $\xi_{0}\sim v_{\mathrm F}/\Delta$ is the 
coherence length in the superconductor.  On the other hand, the regime 
$\ell<\xi_{T}$ belongs to the ballistic regime, if $\ell>d_{\mathrm 
N}$.

To treat the screening problem, a general linear--response formula was 
derived which yields a non--local current--field relation in terms of 
the zero--field Green's functions.  These functions are obtained from 
the solution of the Eilenberger equation including an impurity 
selfenergy.  In general this has to be done numerically, since 
analytical expressions can be derived only in the clean and in the 
dirty limit.  Finally, the Maxwell equations have to be solved.

The chosen geometry is such that the NS interface lies in the $y$--$z$ 
plane.  Then the current in the normal metal has the form
\begin{equation}
 \label{jy}
 j_{y}(x) = -\int\limits_{0}^{d_{\mathrm N}} K(x,x^\prime)
 A_{y}(x^{\prime})dx^{\prime}\;,
\end{equation}
where $A_{y}$ is the vector potential in transverse gauge, if the 
magnetic field is applied in $z$--direction.  The kernel 
$K(x,x^\prime)$ contains an exponential dependence on the distance 
$|x-x^\prime|$.  The range of the kernel is given by $\ell$ in most of 
the temperature and impurity regimes.  For $\ell>\xi_T$ and $\ell \ll 
d_{\mathrm N}\exp(2d_{\mathrm N}/\xi_T)$ that is, at high 
temperatures, it is given by $\xi_T$.  It has thus a strong 
temperature dependence which has to be contrasted to the case of a 
bulk superconductor, where the range of the Pippard kernel is only 
weakly temperature dependent.

The prefactor of the exponential in the kernel, which in general 
depends on coordinates, is related to the local superfluid density.  
It introduces an additional length scale in the problem, the field 
penetration depth $\lambda(x,T)=\lambda_{\mathrm N}f(x,T)$, where 
$\lambda_{\mathrm N}=(4\pi e^2n/m)^{-1/2}$ is a London like length in 
the normal metal and $f(x,T)$ is a function of temperature and 
position.  The interplay between the range of the kernel and the 
superfluid density determines whether the nonlocal form of the current 
field relation is important.  This has strong consequences on the 
screening properties, especially in the case $\ell >d_{\mathrm N}$, 
which is discussed in detail in Ref.\onlinecite{belzig}.

\begin{figure}[tbp]
\includegraphics[width=0.85\linewidth]{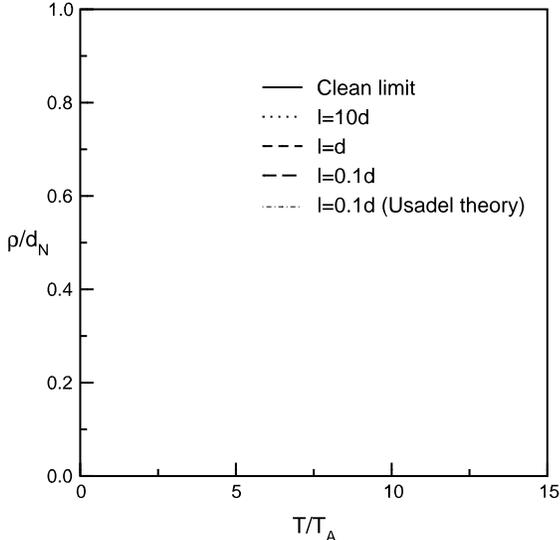}
		\caption{Numerical results of the screening fraction of an NS 
		proximity slab for $\lambda_{\mathrm N}=0.003d_{\mathrm N}$ 
		and different impurity regimes as functions of $T/T_{\mathrm 
		A}$.  The clean limit is given by a solid line and an Usadel 
		theory result for $\ell=0.1d_{\mathrm N}$ is given by a 
		dash-dotted line.  For the intermediate impurity regime three 
		cases are given: $\ell=10d_{\mathrm N}$, $\ell=d_{\mathrm N}$, 
		and $\ell=0.1d_{\mathrm N}$.  }
        \protect\label{figure5}
\end{figure}
In Fig.~\ref{figure5} we reproduce some numerical results\cite{belzig} 
of the screening fraction $\rho/d_{\mathrm N}=-4\pi\chi_{\mathrm N}$ 
in a normal metal layer for $\lambda_{\mathrm N}=0.003d_{\mathrm N}$.  
Five curves as a function of temperature, normalized to the Andreev 
temperature, are given for different mean free paths $\ell$.

For the clean limit, $\ell\to\infty$, at zero temperature, the screening 
can only reach $3/4$ of $-1/4\pi$.  This is due to the infinite range 
of the kernel, or in other words, the complete non--locality of the 
current--field relation.  At a temperature $T\approx 5T_{\mathrm A}$, 
the screening is exponentially reduced by thermal occupation of the 
Andreev levels\cite{stjames}, and at higher temperatures screening is almost 
negligible.

For $\ell=10d_{\mathrm N}$ and $\ell=d_{\mathrm N}$ screening is 
enhanced in comparison to the clean limit at low and high 
temperatures.  At low temperatures screening is more effective for 
$d_{\mathrm N}\leq \ell<\infty$, since the range of the kernel is 
given by finite $\ell$.  At high temperatures, where the range of the 
kernel is $\xi_{T}$, screening is more effective than in the clean 
limit but still nonlocal.  For $\ell=0.1d_{\mathrm N}$, already at 
$T\sim T_{\mathrm A}$ screening is considerably reduced, since the 
superfluid density is suppressed because the mean free path is smaller 
than $d_{\mathrm N}$.


In the diffusive regime, it can be shown that for 
$\ell^2<\lambda_{\mathrm N}d_{\mathrm N}$, the nonlocal relation is 
reduced to the local current--field relation of the Usadel 
theory\cite{usadel}.  In particular, for the case $\lambda_{\mathrm 
N}=0.003d_{\mathrm N}$ corresponding approximately to the experiments 
discussed in this paper, the condition $\ell\ll d_{\mathrm N}$ has to 
be met.  Indeed, in Fig.~\ref{figure5} we illustrate the 
non--negligible difference in screening even for the case 
$\ell=0.1d_{\mathrm N}$, between the local Usadel result and the 
nonlocal approach with elastic scattering described in 
Ref.\onlinecite{belzig}.  For $T\lesssim 5T_{\mathrm A}$ the Usadel 
result exhibits less screening, whereas for $T\gtrsim 5T_{\mathrm A}$ 
it shows a much more pronounced screening tail.

After these recent results, covering a wide range between the clean 
and the dirty limits, we have now the possibility to analyze the 
magnetic response of our relatively clean NS proximity specimens.  To 
compare the experiments with the theory we have used the full form 
(\ref{jy}) valid for all impurity concentrations and solved the 
screening problem for each specimen numerically.

In these theoretical considerations the effect of a rough 
normal metal--vacuum boundary has been neglected completely.  On the 
other hand, at high enough temperatures $T\gg T_{\mathrm A}$, 
where the induced diamagnetism strongly decreases, 
the influence of surface scattering does not play an important role, 
since in this temperature regime the superfluid density at the outer 
boundary is exponentially suppressed.  The screening in this regime 
can be only due to bulk scattering centers and is insensitive to the 
quality of the surface.

The quasiclassical theory for the nonlinear response in the clean 
limit was discussed in detail by Fauch\`ere~\textsl{et 
al.}\cite{fauchere}.  They derived analytical expressions 
for the temperature dependent breakdown field $H_{\mathrm b}(T)$ of a 
clean normal--metal slab of finite thickness in proximity with a bulk 
superconductor in the two limits $T\to 0$ and $T\gg T_{\mathrm 
A}$.  The breakdown field $H_{\mathrm b}$ was determined with a 
Maxwell construction from the magnetization at the bistable regime, 
where two different values of the free energy coexist, characterizing 
a diamagnetic and a field penetration phase.  The spinodals of that 
bistable regime have already been determined in the clean and in the 
dirty limit by Belzig~\textsl{et al.}\cite{belzig:96} numerically.  
They represent the boundaries of superheating and supercooling in the 
first order phase transition.  The thermodynamical magnetic breakdown 
field lies in between the spinodals.  It was determined by the 
intersection of the two asymptotics of the free energy.

For $T\to 0$ the breakdown field saturates at\cite{fauchere} 
\begin{equation}
H_{\mathrm b}(0)\approx \frac{1}{6}\frac{\Phi_{0}}{\lambda_{\mathrm 
N}d_{\mathrm N}}\, ,
\end{equation}
whereas for $T\gg T_{\mathrm A}$ it decays as\cite{fauchere}
\begin{equation}
H_{\mathrm b}(T)\approx \frac{\sqrt{2}}{\pi}\frac{\Phi_{0}}{\lambda_{\mathrm 
N}d_{\mathrm N}}e^{-d_{\mathrm N}/\xi_{T}}\, .
\end{equation}
The temperature dependence is a simple exponential with the exponent 
$d_{\mathrm N}/\xi_{T}=T/T_{\mathrm A}$.  The amplitude of the 
breakdown field was found to scale inversely proportional to the 
normal layer thickness.  Both features are in agreement with 
previous experimental results on $H_{\mathrm b}$ in relatively 
clean AgNb specimens\cite{expsetup}.

In the $H$--$T$ phase diagram a critical temperature 
$T_{\mathrm{crit}}$ was found, below which the first order phase 
transition is observable.  Above $T_{\mathrm{crit}}$ a continuous and 
reversible crossover between the diamagnetic and field penetration 
regime occurs.

Since the breakdown field depends only on thermodynamical 
considerations, it is expected to be less sensitive to the nonlocality 
of the current--field relation than the linear susceptibility.  A 
cylindrical geometry and a barrier at the NS interface is expected to 
change the breakdown field only quantitatively\cite{fauchere}.  On the 
other hand, the influence of scattering centers is reflected in the 
shape of the magnetization curves\cite{belzig:96}.  At present, no 
theoretical results for the breakdown field or magnetization curves 
exist for intermediate impurity concentrations.

\section{Sample preparation and experimental setup}
\label{sec:setup}
The samples we investigated are bundles of cylindrical wires with a 
superconducting core of niobium or tantalum concentrically embedded in 
a normal metal matrix of copper or silver.  The normal metal starting 
materials are of purity 4N to 6N with negligible contribution of 
magnetic impurities.  The purity of the niobium and tantalum starting 
materials was the highest accessible ($RRR\approx~300$), resulting in 
a good ductibility.  We assembled the samples by placing in a close 
fit, a rod of the S material inside a hollow cylinder of the N 
material, with typical outer diameter of about 20\,mm.  The 
cleanliness of the internal N tube surface and the S rod surface is 
very important.  The surfaces of the initial metal pieces were cleaned 
in acid and mechanically smoothed with a special tool in order to 
provide a good metallic contact.  All this was done in an argon 
atmosphere.  The NS cylinder was then embedded in a Cu mantle to 
protect the sample itself from contamination.  After assembly, the 
diameter was reduced mechanically by several steps of swagging and 
co--drawing\cite{flukiger} down to a total diameter of several hundred 
microns.

After etching away the Cu protection, the diameter of the NS specimens 
was further reduced by co--drawing the wires through several diamond 
dies to final values between $15\,\mu\mathrm m$ and $190\,\mu\mathrm 
m$.  The extreme size reduction (by factors of up to 1000) by 
co--drawing resulted in a great enhancement of the quality of the NS 
interface for electronic transmission.  The use of high purity 
starting materials guaranteed a high electronic mean free path.  Some 
of the samples were annealed after the last drawing in an atmosphere 
of argon at a temperature around $650^\circ\mathrm C$ for about one 
hour to remove the effect of cold working in the normal metal.  For 
that purpose the wire of selected size was wound on a small Ag plate.  
As a consequence of the optimized annealing procedure, values of 
$\ell_{\mathrm N}\gtrsim d_{\mathrm N}$ could be achieved.  The mean 
free paths were obtained from resistivity measurements along the 
wires, performed at a temperature $T=10\,\mathrm K$ with a four--point 
method.  The thickness of the normal metal layer as well as the 
surface quality of the wires were determined with the help of SEM 
micrographs.  Several samples were prepared in order to cover a large 
range of parameter space.  We fabricated AgNb samples with a ratio 
$d_{\mathrm N}/r_{\mathrm S}$ of $0.4$ or $1$.  The layer thicknesses 
ranged from $d_{\mathrm N}=2.8\,\mu\mathrm{m}$ to $d_{\mathrm 
N}=28\,\mu\mathrm{m}$ with measured mean free paths ranging from 
$\ell_{\mathrm N}/d_{\mathrm N}=0.12$ to $\ell_{\mathrm N}/d_{\mathrm 
N}=1.3$.  Moreover, we prepared CuNb and CuTa samples, with varying 
ratios of N layer thickness to S core radius, and mean free path 
$\ell$.  This gave us the opportunity to apply the theory described in 
Sec.~\ref{sec:theory} in a wide range of samples.  The values of the 
sample parameters are listed in Table~\ref{tablesamples}.

As a last step, the wires were glued with GE 7031 varnish and thus 
electrically insulated.  After drying, they were removed from the 
silver support, cut to a length of typically 3\,mm to 5\,mm and rolled 
together forming a bundle of 200 to 800 wires.  The wire bundle was 
then placed directly inside the mixing chamber of a dilution 
refrigerator in contact with the liquid 
$\mathrm{^3He}$-$\mathrm{^4He}$ solution, mounted parallel to the 
coil axis of the magnetic ac and dc field.  

The ac magnetic susceptibility was measured at temperatures between 
5\,mK and 7\,K using an rf--SQUID sensor, inductively coupled via a 
dc--flux transformer to the sample. A certain fraction of the voltage 
applied to the
\begin{center}
 \begin{table}[tbp]
	\caption{Table of sample parameters.  
	The measured value $\ell_{\mathrm N}/d_{\mathrm N}$ and the 
	fitting parameter $\ell/d_{\mathrm N}$ are given. In the last 
	column the saturated superheated field at the lowest temperature 
	$H_{\mathrm sh}^{\mathrm sat}$ is shown.}
    \protect\label{tablesamples}
    \begin{tabular}{rccccccccc}
	sample & $d_{\mathrm N}$ & ratio & $T_{\mathrm ann}$ & 
	$\ell_{\mathrm N}$ & $T_{\mathrm A}$ & $\lambda_{\mathrm N}$ & 
	$\ell_{\mathrm N}/d_{\mathrm N}$ & $\ell/d_{\mathrm N}$ & 
	$H_{\mathrm sh}^{\mathrm sat}$\\
	& $[{\mathrm\mu m}]$ & $d_{\mathrm N}/r_{\mathrm S}$ & $[\mathrm 
	^\circ C]$ & $[{\mathrm\mu m}]$ & $[{\mathrm mK}]$ & $[{\mathrm 
	nm}]$ & & & [Oe]\\\hline
	1AgNb & 14.5 & 0.4 & not & 1.8 & 120 & 22   & 0.12 & 0.15 & 1.62\\
	2CuNb & 15.3 & 1.0 & not & 0.9 & 120 & 18.3 & 0.06 & 0.025& 0.42\\
	3AgNb & 5.5  & 0.4 & 700 & 4.3 & 310 & 22   & 0.78 & 0.65 & 15.2\\
	4AgNb & 6.8  & 0.4 & 700 & 5.1 & 250 & 22   & 0.75 & 0.75 & 12.9\\
	5AgNb & 3.3  & 0.4 & 800 & 1.6 & 520 & 22   & 0.49 & 0.55 & 24.7\\
	6AgNb & 28.0 & 0.4 & 550 & 9.0 & 61  & 22   & 0.32 & 0.25 & 2.07\\
	7CuNb & 5.3  & 1.0 & not & 0.9 & 360 & 18.3 & 0.18 & 0.08 &  5.9\\
	12AgNb & 3.6 & 0.4 & not & 1.8 & 480 & 22   & 0.5  & 0.26 & 15.4\\
	14CuTa & 5.0 & 1.0 & not & 1.0 & 380 & 18.3 & 0.2  & 0.07 &  7.0\\
	16AgNb & 3.3 & 0.4 & 800 & 1.7 & 520 & 22   & 0.5  & 0.35 & 24.5\\
	19CuTa & 3.8 & 1.0 & 600 & 2.5 & 510 & 18.3 & 0.67 & 0.13 & 12.9\\
	20AgNb & 2.7 & 0.4 & not & 1.4 & 630 & 22   & 0.53 & 0.24 & 20.2\\
	21AgNb & 3.6 & 0.4 & 800 & 1.0 & 480 & 22   & 0.28 & 0.75 & 15.8\\
	23CuNb & 5.7 & 0.2 & not & 1.4 & 340 & 18.3 & 0.24 & 0.11 &  8.4\\
	24CuNb & 2.5 & 0.2 & not & 0.9 & 760 & 18.3 & 0.35 & 0.15 & 23.9\\
	25CuNb & 1.6 & 0.2 & not & 0.5 & 1200& 18.3 & 0.27 & 0.19 & 40.5\\
	28AgNb & 14.5& 0.4 & 700 & 5.6 & 120 & 22   & 0.39 & 0.45 & 4.89\\
	33AgNb & 7.7 & 1.0 & 700 & 5.3 & 220 & 22   & 0.7  & 0.35 & 10.2\\
	34AgNb & 3.1 & 0.4 & 700 & 4.0 & 540 & 22   & 1.27 & 0.55 & 26.2\\
	35AgNb & 2.8 & 0.4 & 670 & 3.6 & 620 & 22   & 1.3  & 0.55 & 30.8\\
	36AgNb & 5.1 & 1.0 & 670 & 5.4 & 330 & 22   & 1.05 & 0.4  & 14.1\\
	37AgNb & 4.2 & 0.4 & 670 & 3.5 & 410 & 22   & 0.84 & 0.45 & 18.9\\
	38AgNb & 3.8 & 0.4 & not & 1.2 & 450 & 22   & 0.32 & 0.18 & 10.6\\
  \end{tabular}
 \end{table}
\end{center}
\noindent  primary ac--coil was also mutually fed into the flux 
transformer loop, using the SQUID as a null detector.  This allowed us 
to measure the susceptibility of the samples with a relative precision 
of about $10^{-4}$.  Typical ac amplitudes $H_{\mathrm{ac}}$ are 
between $0.06\,\mathrm{mOe}$ and $33\,\mathrm{mOe}$, and frequencies 
$16\,\mathrm{Hz}$, $32\,\mathrm{Hz}$, $80\,\mathrm{Hz}$, and 
$160\,\mathrm{Hz}$.  The temperature was measured through the 
Curie--type magnetic susceptibility of the paramagnetic salt CMN 
(cerium magnesium nitrate), which was calibrated with two Ge 
resistors.  The methods and whole apparatus are explained in more 
detail in Ref.\onlinecite{expsetup}.

\section{Linear magnetic response}
\label{sec:linearresponse}
\subsection{Experimental results}
In the following we report on the temperature dependent 
ac--susceptibility without external dc--field, which we will address 
as linear response.  We show here the results for some typical 
samples, which represent different behaviors and sample parameters, 
namely annealed silver--niobium samples with a large measured mean 
free path $\ell_{\mathrm N}$ as well as not annealed ones with smaller 
$\ell_{\mathrm N}$.  Some of them have a ratio 
$d_{\mathrm N}/r_{\mathrm S}$ of $1$, the normal layer
\begin{figure}[tbp]
\includegraphics{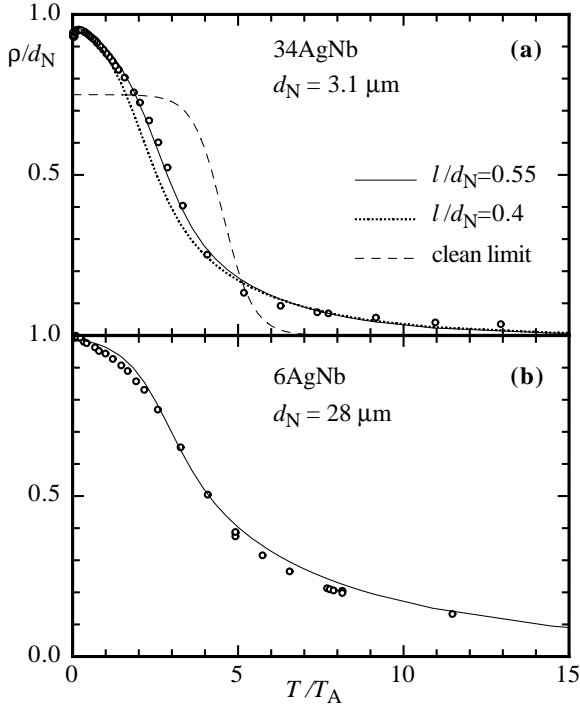}
		\caption{Susceptibility data~($\circ$) and numerical 
		fits~(solid lines) of silver niobium samples as functions of 
		$T/T_{\mathrm A}$.  (a) Fitting parameter $\ell/d_{\mathrm 
		N}=0.55$.  The dashed line denotes the clean limit and the 
		dotted line denotes the $\ell/d_{\mathrm N}=0.4$ curve for 
		comparison.  (b) Fitting parameter $\ell/d_{\mathrm N}=0.25$.  
		}
        \protect\label{fit6/34}
\end{figure}
\begin{figure}[tbp]
\includegraphics{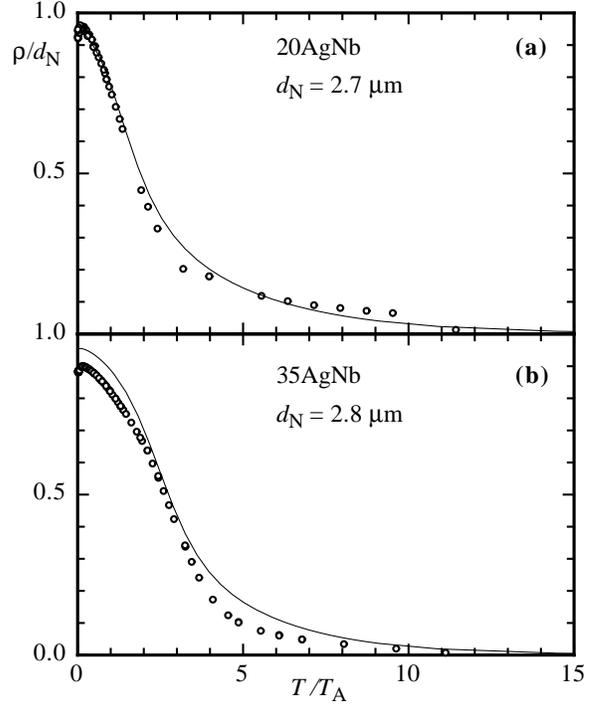}
		\caption{Susceptibility data~($\circ$) and numerical 
		fits~(solid lines) of silver niobium samples as functions of 
		$T/T_{\mathrm A}$.  Fitting parameters (a) $\ell/d_{\mathrm 
		N}=0.25$ and (b) $\ell/d_{\mathrm N}=0.55$.  }
        \protect\label{fit35/20}
\end{figure}
\begin{figure}[tbp]
\includegraphics{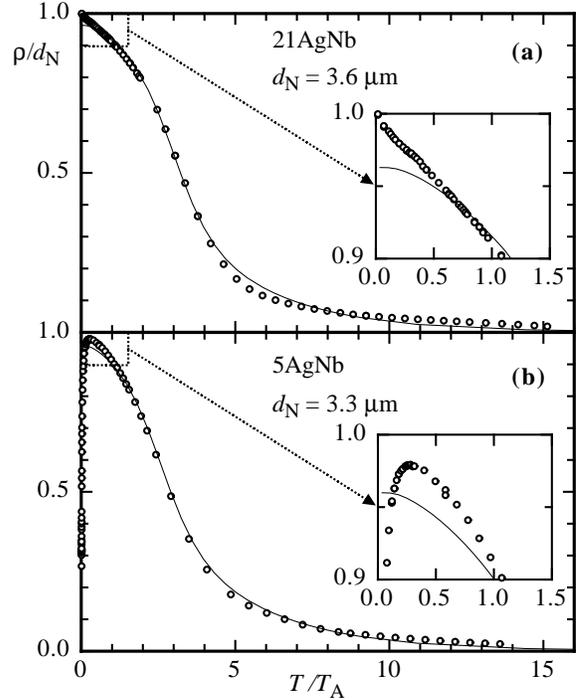}
		\caption{Susceptibility data~($\circ$) and numerical 
		fits~(solid lines) of silver niobium samples as functions of 
		$T/T_{\mathrm A}$.  Fitting parameters (a) $\ell/d_{\mathrm 
		N}=0.75$ and (b) $\ell/d_{\mathrm N}=0.55$.  The insets are 
		zooms in the low temperature region for both samples.  }
        \protect\label{fit21/5}
\end{figure}
\begin{figure}[tbp]
		\includegraphics{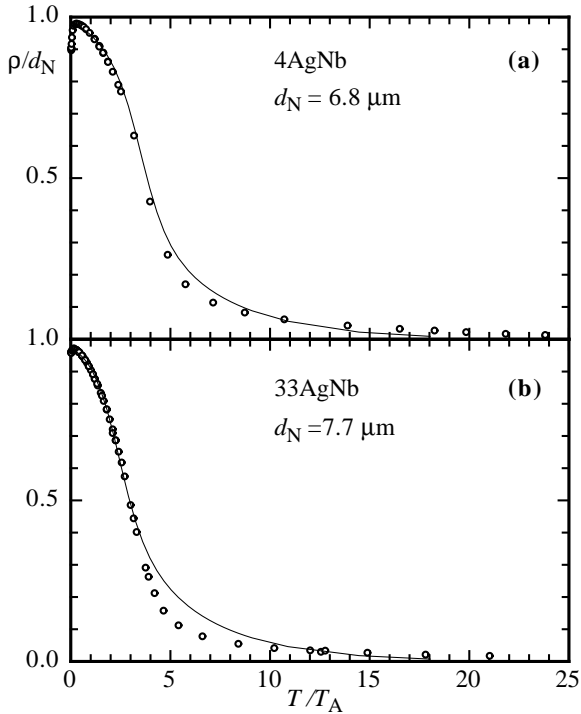} \caption{Susceptibility 
		data~($\circ$) and numerical fits~(solid lines) of silver 
		niobium samples as functions of $T/T_{\mathrm A}$.  (a) ratio 
		$d_{\mathrm N}/r_{\mathrm S}=0.4$ and fitting parameter 
		$\ell/d_{\mathrm N}=0.75$.  (b) ratio $d_{\mathrm 
		N}/r_{\mathrm S}=1$ and fitting parameter $\ell/d_{\mathrm 
		N}=0.4$.  The fit for sample~33AgNb shows a bigger deviation 
		from the data for $5\lesssim T/T_{\mathrm A}\lesssim 15$ than 
		the fit for sample~4AgNb.  }
        \protect\label{fit4/33}
\end{figure}
\noindent having a rather strong curvature, while others have a ratio 
$0.4$ and therefore an almost flat normal layer.  Also shown are not 
annealed copper--niobium samples with ratios of $1$ and $0.2$ and a 
relatively smaller mean free path $\ell_{\mathrm N}$ as well as two 
copper--tantalum samples with a ratio $d_{\mathrm N}/r_{\mathrm S}=1$.

\begin{figure}[tbp]
\includegraphics{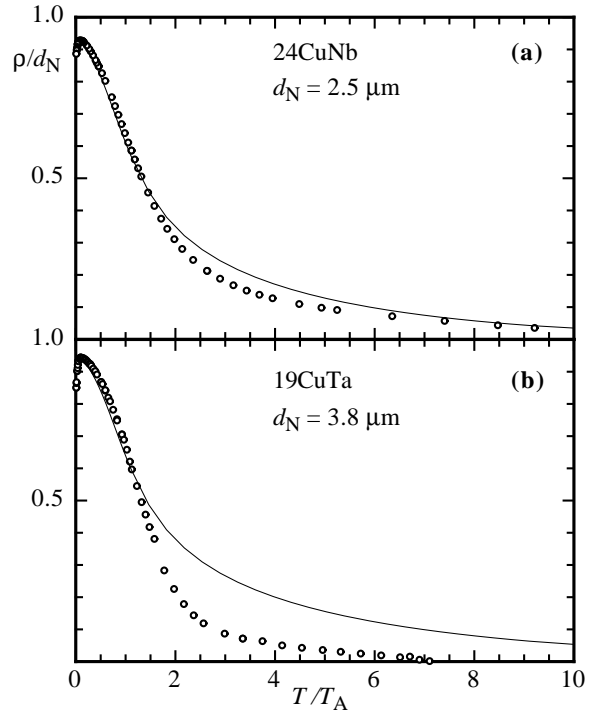}
		\caption{Susceptibility data~($\circ$) and numerical 
		fits~(solid lines) of (a) the copper niobium sample~24CuNb, 
		fitting parameter $\ell/d_{\mathrm N}=0.15$ and (b) the copper 
		tantalum sample~19CuTa, fitting parameter $\ell/d_{\mathrm 
		N}=0.13$ as functions of $T/T_{\mathrm A}$.  }
        \protect\label{fit19/24}
\end{figure}

The data of the diamagnetic screening fraction $\rho(T)/d_{\mathrm N}$ 
of the N layer are shown in 
Figs.~\ref{fit6/34},~\ref{fit35/20},~\ref{fit21/5},~\ref{fit4/33}, 
and~\ref{fit19/24} as a function of temperature in units of the 
Andreev temperature $T_{\mathrm A}$ for each sample.  The rather 
strong development of the induced superconductivity in the normal 
metal right below the transition temperature $T_{\mathrm c}$ of the 
superconductor can be observed quite clearly.

From the susceptibility $\chi(T)$ measured in arbitra\-ry units the 
screening fraction was obtained as
\[
\rho(T)/d_{\mathrm N}=r_{\mathrm
S}/ d_{\mathrm N}\cdot [(1+\chi_{\mathrm N}(T)/\Delta\chi_{\mathrm
S}(T_{\mathrm c}))^{1/2}-1]\,.
\]
Here
\[
\chi_{\mathrm
N}(T)/\Delta\chi_{\mathrm S}(T_{\mathrm
c})=(\chi(T)-\chi_{0})/(\chi_{0}-\chi_{\infty})
\]
is the fraction of the temperature dependent susceptibility of the 
normal metal with respect to the total diamagnetic transition of the 
superconductor at its critical temperature $T_{\mathrm c}$. The exact 
height of both $\Delta\chi_{\mathrm S}$ and $\chi_{\mathrm N}(T)$ 
depends on the value of the susceptibility $\chi_{0}$ right below 
the transition of the superconductor.  Since that transition for the 
niobium samples is outside of our accessible 
\noindent temperature range, values at the highest temperatures were 
used for extrapolation.  The susceptibility above the
superconducting transition $\chi_{\infty}$ 
was obtained from background measurements.  Through the ratio of the 
radius of the superconductor $r_{\mathrm S}$ to the normal layer 
thickness $d_{\mathrm N}$ the normal layer curvature of the composite 
cylindrical sample is taken into account.

In a local picture as given by the Orsay Group on 
Superconductivity\cite{degennes} the screening fraction 
$\rho/d_{\mathrm N}$ represents the screening length normalized to the normal 
layer thickness, with $\rho$ the thickness of the part of the normal 
metal, out of which the magnetic flux is screened.  For our 
samples that picture is not valid, since the current--field relation 
is nonlocal.  Here the screening fraction $\rho/d_{\mathrm N}$ 
represents the susceptibility of N, as related to a model of a one 
dimensional system.

\subsection{Discussion of the fits}
The temperature dependence of the diamagnetic susceptibility of these 
samples is neither accounted for by the clean limit nor the dirty limit 
theory.  The values of the measured mean free path $\ell_{\mathrm N}$ 
indicate that the samples are in an intermediate impurity regime.  
They are not diffusive enough for the dirty limit, because 
$\ell_{\mathrm N}$ is of the order of $d_{\mathrm N}$ for most of the 
samples.  However, since they have a small density of scattering 
centers, they are not in the clean limit, either.  From here on we 
will address them as relatively clean.

We fitted the experimental data of the screening fraction with the 
theoretical results obtained for a one-dimensional geometry, as 
described in Sec.~\ref{sec:theory}.  The fits are shown in 
Figs.~\ref{fit6/34},~\ref{fit35/20},~\ref{fit21/5},~\ref{fit4/33}, 
and~\ref{fit19/24}.

The parameters $d_{\mathrm N}$ and $\lambda_{\mathrm N}$ entering the 
numerical calculations were obtained from measurements and well known 
material constants.  The only fitting parameter was then the mean free 
path $\ell$, which enters into the theory through the fraction 
$\ell/d_{\mathrm N}$.  The fits were performed giving the experimental 
data at an intermediate temperature regime the most weight, where the 
increase of the susceptibility was the steepest.  To illustrate the 
sensitivity of the fits to the value of the fit parameter 
$\ell/d_{\mathrm N}$, in Fig.~\ref{fit6/34}(a), we give in addition to 
the fitted curve with $\ell/d_{\mathrm N}=0.55$, the curve for 
$\ell/d_{\mathrm N}=0.4$.

The curves obtained from the newly developed 
approach with arbitrary scattering center concentration fit the 
experimental data well over the whole temperature range.  
The corresponding clean limit curve does not fit the data, as 
shown for comparison in Fig.~\ref{fit6/34}(a).

The values of the fitting parameter $\ell/d_{\mathrm N}$ reproduce the 
measured values $\ell_{\mathrm N}/d_{\mathrm N}$ rather well.  It has 
to be emphasized, that the fits are rather good, if one takes into 
account the difference in geometry between experiment and theory, the 
neglect of the boundary roughness in the theory and other 
imperfections inevitably present in the samples.  This means, that the 
quasiclassical theory of the proximity effect with a finite mean free 
path parameter $\ell$ due to a low concentration of elastic 
scatterers\cite{belzig}, is now able to explain the linear 
susceptibility data of our relatively clean NS specimens.  A list of 
all the samples shown here is given in Table~\ref{tablesamples}.  
Additionally more samples are listed together with the mean free path 
$\ell$ which gave the best fit.

For the silver--niobium samples with a measured mean free path 
$\ell_{\mathrm N}/d_{\mathrm N}\approx 1$ the fitting parameter 
$\ell/d_{\mathrm N}$ lies between $0.4$ and $0.8$.  The measured mean 
free path of AgNb samples with $\ell_{\mathrm N}/d_{\mathrm N}<0.3$ is 
also reproduced quite well in the fits.  A typical example is the 
silver--niobium sample~6AgNb shown in Fig.~\ref{fit6/34}(b).  It is 
suprising, that for most of the samples the agreement between the only 
fitting parameter $\ell/d_{\mathrm N}$ and the measured $\ell_{\mathrm 
N}/d_{\mathrm N}$ is so good.  However, the parameter $\ell$ has to be 
viewed as an effective mean free path, which contains the scattering 
from bulk impurities as well as from the surface.  The measured mean 
free path $\ell_{\mathrm N}$ is also an effective quantity, which 
contains surface effects, but it is yielded from transport 
measurements along the axis of the wires.  There the surface is 
expected to influence the mean free path in a different way, because 
the geometry of the relevant trajectories differs from the ones in the 
proximity case.

In the following, we describe as an example, the quality of the fits 
for the silver--niobium samples.  These samples show the best 
agreement between the measured magnetic susceptibility and the theory.  
The results for some of them are illustrated in 
Figs.~\ref{fit6/34},~\ref{fit35/20},~\ref{fit21/5}, and~\ref{fit4/33}.  
In general, the fits given by the solid lines, describe the 
experimental data well.  In particular, for some of these samples we 
observe considerable deviations at the lowest temperatures 
($T<T_{\mathrm A}$) where the theoretical curves saturate as shown in 
Fig.~\ref{figure5}.  Moreover, the deviations are different for 
nominally similar samples.  For example, sample~21AgNb (inset of 
Fig.~\ref{fit21/5}(a)) show higher $\chi_{\mathrm N}$ values than the 
theory, while sample~5AgNb (inset of Fig.~\ref{fit21/5}(b)) shows also 
higher values and in addition a strong reentrance of $\chi_{\mathrm 
N}$ at the lowest temperatures.  These apparently contradictory 
results are possibly related to the surface quality.  At this point, 
it is not clear how a non--ideal reflecting surface affects the 
susceptibility, neither experimentally nor theoretically.  In 
addition, silver--niobium samples with roughly the same $d_{\mathrm 
N}$ and $\ell$ show strongly different levels of reentrance.  The 
origin of these different behaviors will be subject of further 
investigations.

The fit shown in Fig.~\ref{fit35/20}(b) is rather poor, which could be 
partially explained by the geometrical non--uniformity of 
sample~35AgNb, showing effectively two different values of $d_{\mathrm 
N}$ (see also Fig.~\ref{MH} and Sec.~\ref{sec:nonlinear}).


For $T\gtrsim T_{\mathrm A}$ the fits of the AgNb specimens are quite 
good.  For all the specimens, in a medium temperature regime, the 
theoretical susceptibility is a bit too high with respect to the 
measured one.  This behavior can be observed more clearly for the 
samples in Fig.~\ref{fit4/33}.  Their normal layer thickness with 
values of $d_{\mathrm N}=6.8\,\mu\mathrm{m}$ and $d_{\mathrm 
N}=7.7\,\mu\mathrm{m}$ as well as the measured mean free path with 
values of $\ell_{\mathrm N}=5.1\,\mu\mathrm{m}$ and $\ell_{\mathrm 
N}=5.3\,\mu\mathrm{m}$ are approximately the same.  The only 
difference between the two samples is the ratio $d_{\mathrm 
N}/r_{\mathrm S}$, which for sample~4AgNb is $0.4$ and for 
sample~33AgNb is $1$.  As expected, the sample~33AgNb with the bigger 
ratio and therefore with the stronger curvature of the normal layer 
shows the more pronounced deviation in the medium temperature regime.

The susceptibility of the copper--niobium samples is also described by 
the theory rather well.  A typical example is shown in 
Fig.~\ref{fit19/24}(a).  For similar samples the fitted values of 
$\ell/d_{\mathrm N}$ are very small, lying between $0.02$ and $0.2$ 
(see Table~\ref{tablesamples}).  In contrast to the silver--niobium 
samples with very small mean free paths, for the copper--niobium 
samples the value of $\ell/d_{\mathrm N}$ is only about half of the 
measured value of $\ell_{\mathrm N}/d_{\mathrm N}$.  The CuNb samples 
show similar deviations in the different temperature regimes as the 
silver--niobium samples.

The poorest agreement between experiment and theory is met for the 
copper--tantalum samples.  An example is shown in 
Fig.~\ref{fit19/24}(b).  As the temperature is increased the 
theoretical curve starts to show much more screening than the 
experiment.  The deviation starts to show up, where the induced 
diamagnetism is about half reduced.  At higher temperatures the 
difference between the two curves is even larger.  Moreover, the 
gained fitting parameter $\ell/d_{\mathrm N}$ is much smaller than the 
measured $\ell_{\mathrm N}/d_{\mathrm N}$.  Interestingly, the two 
copper--tantalum samples investigated (14CuTa and 19CuTa) show the 
same strong disagreement between $\ell$ and $\ell_N$.  This suggests 
that some other effects should be considered for the combination of 
these two materials.

\section{Nonlinear response measurements}
\label{sec:nonlinear}
We have investigated for our NS proximity samples the nonlinear 
magnetic response.  With the experimental setup described in 
Ref.\onlinecite{expsetup} ac--susceptibility measurements at 
constant temperature as a function of a dc magnetic field were 
performed, as well as isothermal dc--magnetization curves.
\begin{figure}[tbp]
\includegraphics{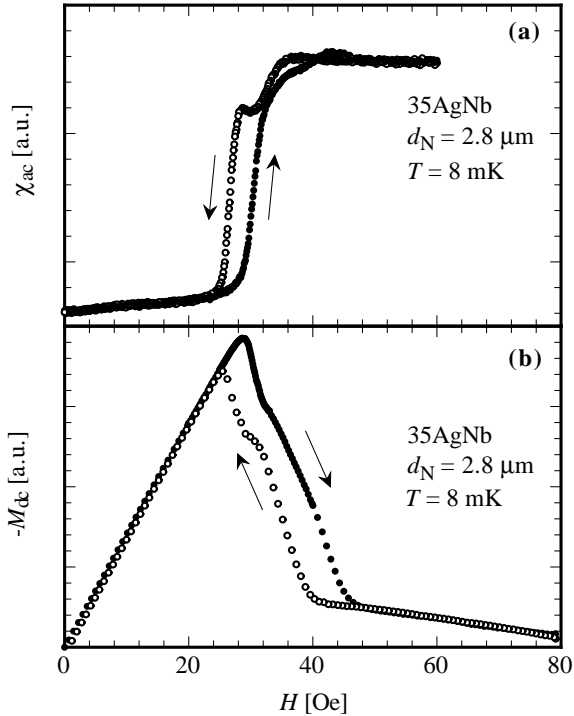}
		\caption{(a) Nonlinear ac susceptibility as a function of 
		magnetic field $H$.  (b) Isothermal dc magnetization curve.  }
        \protect\label{MH}
\end{figure}

Fig.~\ref{MH} shows the nonlinear susceptibility and the 
dc--magnetization curve for one of the cleanest silver--niobium 
samples at $T= 8\,\mathrm{mK}$.  The magnetic breakdown of the induced 
superconductivity occurs at a magnetic field $H_{\mathrm b}\approx 
25\,\mathrm{Oe}$.  Below the breakdown field the magnetic flux is 
totally screened, and above, it enters into the normal metal layer.  
The first order phase transition shows the features of superheating at 
increasing field and of supercooling at decreasing field.  That kind 
of hysteretic behavior was already observed and discussed in the dirty 
limit by the Orsay Group on Superconductivity\cite{degennes}.  This 
particular sample shows a small second transition, which is due to the 
niobium core not being perfectly centered resulting in effectively two 
slightly different normal layer thicknesses.  We notice that the 
magnetic breakdown transition is not infinitely sharp.  In our samples 
consisting of several hundred wires the superheated and supercooled 
transitions are triggered for each individual wire at slightly 
different fields, leading to a statistical broadening of the breakdown 
jumps.

From the isothermal $\chi_{\mathrm N}(H)$--curves we have determined 
the supercooled and superheated fields as the fields at the middle of 
each transition.  In this way values of $H_{\mathrm sc}$ and 
$H_{\mathrm sh}$ at different temperatures were obtained.

\begin{figure}[tbp]
\includegraphics{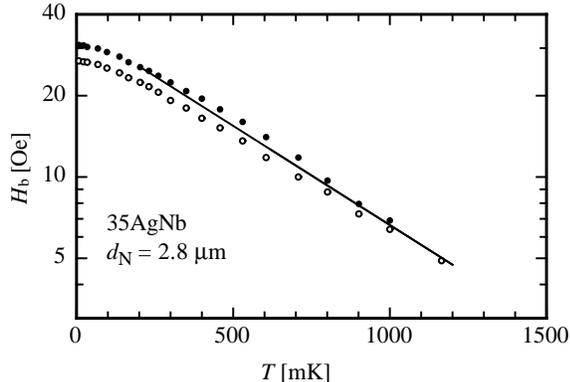} 
				\caption{Measured 
				supercooled ($\circ$) and superheated ($\bullet$) 
				breakdown fields of the annealed silver--niobium 
				sample~35AgNb with $d_{\mathrm N}=2.8\,\mu\mathrm{m}$.  
				The line represents the clean limit theory of a one 
				dimensional NS slab with a correction factor $0.3$.  }
        \protect\label{HbT}
\end{figure}

For our relatively clean silver--niobium samples at temperatures 
higher than the Andreev temperature $T_{\mathrm A}\propto v_{\mathrm 
F}/d_{\mathrm N}$, the experimental breakdown fields follow the 
exponential dependence $H_{\mathrm b}(T)\propto 1/d_{\mathrm 
N}\exp(-d_{\mathrm N}/\xi_{\mathrm N}(T))$.  The experimentally found 
values of $\xi_{\mathrm N}(T)$ reproduce the theoretical clean limit 
coherence length in silver $\xi_{T}=\hbar v_{\mathrm F}/2\pi 
k_{\mathrm B} T=1.69\,\mu\mathrm{m}/T(\mathrm{K})$ within a few 
percent.  For the sample~35AgNb this is illustrated in Fig.~\ref{HbT}, 
where the experimental supercooled and superheated breakdown fields as 
a function of temperature are given.  The theoretical clean limit 
breakdown field of a one dimensional NS slab for $T\gg T_{\mathrm A}$ 
is shown as a solid line.  A factor of about $0.3$ has been used to 
shift down the theoretical curve to fit the experiment.  We notice 
that the temperature dependence of the breakdown field agrees 
qualitatively with the clean limit result\cite{fauchere}, obtained for 
a one dimensional NS slab assuming ideal boundary conditions, e.g.  
ideal normal electron transmission at the NS interface.  
Quantitatively, the measured breakdown fields are a factor of about 
$0.3$ smaller than the high temperature theoretical curve in 
Fig.~\ref{HbT}.

This discrepancy could have different reasons.  First, in our samples 
normal reflections at the NS interface may happen due to the mismatch 
of the Fermi velocities between N and S and impurities or 
interdiffusion.  Second, our samples have a cylindrical geometry, 
which weakens the proximity effect with respect to the one dimensional 
flat geometry considered in the theory.  And third, the
influence of 
impurity scattering on the degree of nonlocality is expected to lead 
to quantitative deviations of the breakdown field data from 
the clean limit theory.

\begin{figure}[tbp]
\includegraphics{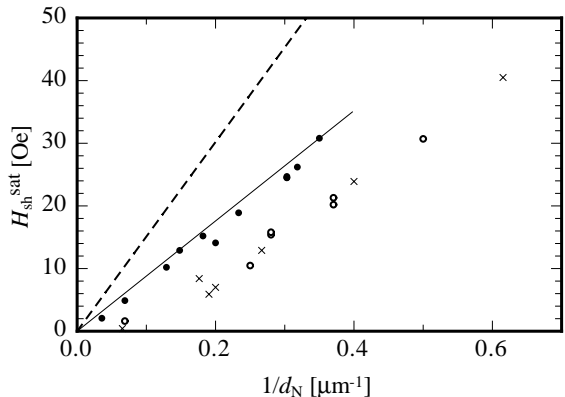} 
				\caption{$1/d_{\mathrm N}$--dependence of the 
				saturated superheated field for silver--niobium samples with 
				$\ell_{\mathrm N}/d_{\mathrm N}\approx 1$ ($\bullet$).  
				Slope of solid line: $88\,\mathrm{Oe\mu m}$.  The 
				dashed line represents the clean limit result, with 
				slope: $150\,\mathrm{Oe\mu m}$.  Also shown are less 
				clean silver--niobium samples ($\ell_{\mathrm 
				N}/d_{\mathrm N}<0.3$,$\circ$) and copper samples 
				($\times$), which we do not consider in the fit.  }
        \protect\label{Hsh(1/dN)}
\end{figure}

In Fig.~\ref{Hsh(1/dN)} the saturated superheated fields of the 
silver--niobium samples are plotted versus the inverse normal layer 
thickness $1/d_{\mathrm N}$.  The samples with $\ell_{\mathrm 
N}/d_{\mathrm N}\approx 1$ and $d_{\mathrm N}/r_{\mathrm S}=0.4$ show 
a clear $1/d_{\mathrm N}$--dependence of the breakdown field, in 
agreement with the clean limit theory\cite{fauchere}.  On the other 
hand, the 
absolute values of the saturated breakdown fields are about a factor 
of $1.7$ lower than in the clean limit theory.  All the other samples 
with a ratio $d_{\mathrm N}/r_{\mathrm S}=1$ or with a lower mean free 
path $\ell_{\mathrm N}$ show smaller breakdown fields.  As already 
mentioned, for the copper samples the influence of a reduced normal 
electron transmission coefficient at the NS boundary could be 
especially strong leading to a further suppression of the breakdown 
field.

The silver--niobium samples with $\ell_{\mathrm N}/d_{\mathrm N}<0.3$ 
and the copper samples also show a temperature dependence $H_{\mathrm 
b}(T)\propto \exp(-d_{\mathrm N}/\xi_{\mathrm N}(T))$, but for these 
samples experimentally we found $\xi_{\mathrm N}(T)=p\cdot\xi_{T}$, 
where the prefactor $p$ is about $0.3$ to $0.6$\cite{expsetup}.  The 
breakdown field data of these samples can no longer be explained by 
the clean limit theory, since the impurity scattering is too strong.

Although the temperature dependence of the breakdown field is the same 
as given by the clean limit, the full form of the magnetization curves 
$M(H)$ of the cleanest AgNb samples can not be explained within this 
limit.  The slope below the transition does not correspond to the 
$-3/4$ of a perfect diamagnet and additionally, the observed finite 
screening tail above the transition (see Fig.~\ref{MH}(b)) does not 
appear in the clean limit.  Clearly, agreement between the 
quasiclassical theory and the magnetization data could be achieved by 
taking into account the finite mean free path $\ell$.  (Note the 
similarity between the magnetization curve in Fig.~\ref{MH}(b) and the 
theoretical dirty limit magnetization curves from 
Ref.\onlinecite{belzig:96}, which have the right slope as well as 
a finite screening tail, but give the wrong field scale.)  Theoretical 
work on $M(H)$ for arbitrary low impurity concentrations is still 
needed.

\section{Conclusions}
We investigated the magnetic response of NS proximity layers of high 
purity, and compared the experimental data with the quasiclassical 
theory.  A large number of samples containing combinations of Ag and 
Cu with Nb and Ta has been fabricated and measured.  All the samples 
show an induced diamagnetic susceptibility close to $-1/4\pi$ for 
temperatures $T\leq T_{\mathrm A}$.  The mean free path $\ell_{\mathrm 
N}$ obtained from resistivity measurements varied between 
$0.1d_{\mathrm N}$ and $1.3d_{\mathrm N}$.  With the results of 
Belzig~\textsl{et al.} which are based on the quasiclassical theory 
including arbitrary impurity concentrations, we were able to reproduce 
the experimental data quite well, using a mean free path $\ell$ as the 
only fitting parameter.  The mean free path $\ell$ obtained in this 
way agrees within a factor of 2 with the mean free path $l_{\mathrm 
N}$ determined by resistivity measurements.  This good agreement shows 
that the linear diamagnetic response of a proximity system with 
arbitrary impurity concentration is very well described with the 
present nonlocal approach.  However, the very low temperature 
behavior, where some deviations occur and in addition an unexplained 
reentrance of the susceptibility appears, is still not understood.

Magnetization measurements show a first order transition at a 
breakdown field from a state with almost complete flux expulsion to a state 
with weak screening.  The temperature dependence of the breakdown 
field for the cleanest specimens is in accordance with the clean 
limit.  Nevertheless, the complete magnetization curve, in agreement 
with the linear susceptibility data, reflects the finite mean free 
path of the samples.

With this work we show that the proximity theory, based on the 
quasiclassical approximation including the full nonlocal current 
response, can successfully describe the magnetic response of very 
clean samples.  The experimentally unavoidable low level of impurities 
can be accounted for with a mean free path $\ell$ which is of the 
order of the sample size.

\section*{Acknowledgments}
We would like to thank G.~Blatter, C.~Bruder, A.~Fauch\`ere, 
G.~Sch{\"o}n, and A.~Zaikin for helpful discussions.  We acknowledge 
partial support from the ``Schweizerischer Nationalfonds zur 
F{\"o}rderung der Wissen\-schaft\-lichen Forschung'' and the 
``Bundesamt f{\"u}r Bildung und Wissenschaft'' (EU Program ``Human 
Capital and Mobility'').  W.~B. likes to thank the ETH Z{\"u}rich for 
hospitality and the Deutsche Forschungsgemeinschaft (grant 
No.~Br1424/2-1) for financial support.

\end{multicols}

\begin{thebibliography}{00}
\bibitem{flukiger} {We kindly thank Prof.~Fl{\"u}kiger
at the University of Geneva for valuable suggestions and for 
providing us his facilities for sample preparation.
}
\end{thebibliography}
\end{document}